\documentclass[journal]{IEEEtran}

\usepackage{graphicx}
\usepackage{epsf}
\usepackage{mathtools}
\usepackage{amssymb}
\usepackage{amsmath, amsfonts}
\usepackage{latexsym}
\usepackage{multirow}
\usepackage{multicol}
\usepackage[table]{xcolor}
\usepackage{color}

\usepackage{tabularx}
\usepackage{lipsum}

\usepackage{algorithm}
\usepackage{algorithmic}

\usepackage{mathrsfs}

\usepackage{fixltx2e}

\usepackage[mathscr]{euscript}

\usepackage{commath}
\usepackage{MnSymbol}

\usepackage{subcaption}

\usepackage{mathtools} 
\DeclarePairedDelimiterX\setc[2]{[}{]}{\,#1 \;\delimsize\vert\; #2\,}
\DeclarePairedDelimiterX\parth[2]{(}{)}{\,#1 \;\delimsize\vert\; #2\,}

\PassOptionsToPackage{hyphens}{url}
\usepackage[unicode=true, bookmarks=true, bookmarksnumbered=true, bookmarksopen=true, bookmarksopenlevel=1, breaklinks=false, pdfborder={0 0 0}, pdfborderstyle={}, backref=false, colorlinks=false]{hyperref}

\usepackage{theorem}
{
\theorembodyfont{\rmfamily}

}

\usepackage{soul}

\definecolor{orange}{RGB}{255,127,0}
\definecolor{blue}{RGB}{0,0,255}
\definecolor{red}{RGB}{255,0,0}
\definecolor{green}{RGB}{50,160,50}
\definecolor{grey}{RGB}{125,120,125}
\definecolor{purple}{RGB}{125,0,125}

\newcommand{\MK}[1]{{\color{black}{#1}}}

\usepackage{hyperref}
\hypersetup{
  colorlinks   = true, 
  urlcolor     = black, 
  linkcolor    = black, 
  citecolor   = black 
}

\begin{document}
{
\title{{\fontsize{20}{2}\selectfont Cross-Layer Performance Evaluation of C-V2X}}

\author
{
Dhruba Sunuwar and Seungmo Kim, \textit{Senior Member}, \textit{IEEE}


\thanks{D. Sunuwar and S. Kim are with the Department of Electrical and Computer Engineering, Georgia Southern University, Statesboro, GA, USA. This work is supported by the Georgia Department of Transportation (GDOT) via grant RP21-08.}
}

\maketitle
\begin{abstract}
\MK{As self-driving cars increasingly penetrate our daily lives, vehicle-to-everything (V2X) communications are emerging as one of the key enabler technologies. However, the dynamicity of vehicles (one of whose causes is the mobility of vehicles) often complicates it even further to evaluate the performance of a V2X system. We have been building a system-level simulator dedicated to assessing the performance of V2X communications. We highlight that the simulator features the incorporation of (i) intelligent transportation system (ITS) scenarios in geographical setup and (ii) physical (PHY) and radio resource control (RRC) cross-layer performance evaluation capability. In particular, this abstract reports the status of our implementation of the modulation and coding scheme (MCS).}
\end{abstract}


\section{Introduction}\label{sec_intro}
\subsubsection{Background}\label{sec_intro_background}
\MK{
Vehicle-to-Everything (V2X) communications have the potential to significantly reduce the number of accidents and associated fatalities \cite{intro1}. This capability positions V2X communications at the core of intelligent transportation systems (ITS) for connected vehicle environments.

Nonetheless, in November 2020, the U.S. Federal Communications Commission (FCC) issued a rule repurposing the 5.850-5.895 GHz spectrum (the ``lower 45 MHz'') of the 5.9 GHz band for the expansion of unlicensed communications (e.g., Wi-Fi), only leaving the 5.895-5.925 GHz spectrum (the ``upper 30 MHz'') for ITS operations \cite{rns23_1}.

Within the remaining upper 30 MHz, the FCC has begun endorsing cellular-V2X (C-V2X) through waivers granted to state departments of transportation (DOTs) and other stakeholders \cite{rns23_2}\cite{rns23_3}. As such, we regard this an urgent situation for a thorough investigation of C-V2X performance.
}

\subsubsection{Related Work}\label{sec_intro_related}
\MK{
C-V2X has already formed a massive body of literature on many fronts including performance measurement metric \cite{irt_06}\cite{vtc23_dhruba} and performance evaluation methods based on theoretical/mathematical analysis \cite{tvt22}\cite{access20_bjkim}, computer simulations \cite{arxiv2208}\cite{vtc23_zach}, channel sounding \cite{nyu}, applications \cite{access19}-\cite{tiv23}.

The prior art certainly provides profound insights, yet is not directly conclusive whether the reduced 30-MHz bandwidth makes it feasible to operate C-V2X on realistic road and traffic scenarios. The same limitation can be found in the current literature of V2X safety-critical applications \cite{ZoK22a}: the proposals lay out approaches to support such applications but leave it unaddressed what the influence will be after the C-V2X got deprived of 60\% of its bandwidth.

The latest literature showed several techniques as a means to resolve the bandwidth constraint. A reinforcement learning-based mechanism was proposed \cite{vtc22_bjkim} to optimize the C-V2X performance according to the road danger level. Moreover, other spectrum bands were considered to support the increasing demand of ITS services--viz., the 4.9 GHz public safety band \cite{iceic22_faizan} and the 5 GHz unlicensed band \cite{arxiv20_bjkim}.
}

\subsubsection{Contributions}
\MK{Addressing the limitations of the aforementioned prior work, this research highlights the following contributions:
\begin{itemize}
\item Building a simulator that features ``vertical integration'' of (i) ITS scenarios in geographic/traffic setup and (ii) PHY/RRC cross-layer performance evaluation capability
\item Implementation of modulation and coding scheme (MCS), resulting in precise C-V2X performance evaluation according to geographic/traffic scenario
\end{itemize}
}

\section{System Model}\label{sec_model}
\subsection{Spatial Setup}\label{sec_model_spatial}
A two-dimensional urban scenario \cite{J3161} with dimensions of 240 m $\times$ 520 m is established in MATLAB. The setup consists of three two-way road segments, with two junctions each comprising an RSU with a range of 150 m and utilizing a 20 MHz bandwidth. Each of the road segments is divided into two directions and each direction consists of two lanes. A total of six directions are considered:  South-North, North-South, East-West 1, West-East 1, East-West 2, and West-East 2. Two parameters, $\lambda$ and $\theta$, correspondingly indicate the densities of vehicles and trucks, which are deployed randomly.

\MK{Notice that the trucks act as physical obstacles blocking V2X connections. The simulator features the ability to visually distinguish connected and blocked links between the RSU and neighboring vehicles. The blocked links are contributed by two types of physical obstacles: buildings and trucks}, which can significantly attenuate the signals transmitted from the RSU to the neighboring vehicles. 

\subsection{LTE-V2X Simulator}\label{sec_model_LTE_V2X_Simulator}
The 3rd Generation Partnership Project (3GPP) Release 15 Long Term Evolution (LTE)-V2X for the \MK{physical-layer (PHY) \cite{3GPP1} and radio resource control (RRC)} functions \cite{3GPP2} has been adopted in the LTE-V2X simulator using MATLAB. The \MK{sensing-based semi-persistent scheduling (SPS)} algorithm for mode 4 has been employed, where the \MK{resource reservation interval} (RRI) is set to 100 msec, and a random \MK{resource reselection counter} (RC) between 5 and 15 is assigned to each vehicle.

The RC \MK{is decremented by 1} after each transmission, and once the RC reaches zero, the vehicle opts for new resources. \MK{A 20-MHz channel} is divided into subchannels in the frequency domain and subframes in the time domain. Subchannels are further divided into resource blocks (RBs), each 180 kHz, and subframes into slots, each of which is 0.5 msec. The number of RBs per subchannel is variable depending on the MCS index. The Urban Micro (UMi)-Street Canyon path loss model has been incorporated in the simulation, considering the city road environment. 

The number of allowed retransmissions for mode 4 sidelink is 1, and the number of subchannels per slot can vary from 1 to 10 \cite{Bazzi}. The congestion control mechanism \cite{ETSI} is integrated, where the \MK{channel busy ratio (CBR) and channel occupancy ratio} (CR) are calculated. When the calculated CR exceeds the CR limit, the vehicle must decrease its CR below the limit. \MK{However, the specific technique to achieve this has not been standardized;} it is left to the implementers to choose any of the following techniques: (a) drop packet transmission/re-transmission, (b) adapt the MCS, or (c) adapt transmission power \cite{Mans}. \textit{We highlight that this work is complete throughout (a) and (c), which provides researchers the ability to precisely measure the performance of a C-V2X system that is applied to real-world scenarios.}

\subsubsection{Performance Metrics}\label{sec_model_performance_metrics}
The \textit{end-to-end latency} and \MK{\textit{packet delivery rate (PDR)}} serve as the key performance metrics for the LTE-V2X system. Both metrics are calculated for vehicles within the range of the RSU.

The PDR is calculated using \MK{Eq. (\ref{eq_pdf})}.

\begin{align}\label{eq_pdf}
\text{PDR} = \frac{\# \text{ Transmitted and successfully received packets}}{\# \text{ Transmitted packets} + \# \text{ Dropped packets}}
\end{align}

Since we are focusing on mode 4 of C-V2X, the latency is implemented as the length of time taken from the generation of a message at an RSU to its reception by a vehicle.

\section{Experimental Results}\label{sec_results}
The simulation was run with the following parameters: $\lambda$ = 20 (120 vehicles in total); Tx power = 23 dBm; Rx sensitivity = -97.28 dBm; Message Tx rate = 10 Hz; and inter-broadcast interval = 100 msec. The MCS indices were set to \{7, 11\} with the number of subchannels equal to 2 and 7, respectively.

Fig. \MK{\ref{fig_latency} demonstrates the probability density function (pdf), $f_{X}(x)$}, of the end-to-end latency $x$ (in msec) from a RSU to a vehicle. It is evident that a higher MCS index of 11, which uses a larger number of subchannels, yields an improved latency for the vehicles compared to a lower MCS index of 7.

Although higher MCS indices are expected to offer higher data rates, the transmitted messages become more susceptible to errors due to distance from the RSU, interference, and obstacles. Recall that in the simulation, big trailer trucks and buildings are the major obstacles that result in the reduction of the overall PDR as the MCS index changes from 7 to 11.

\MK{Fig. \ref{fig_PDR} demonstrates the result of PDR: it lays out a cumulative distribution (CDF) for PDR. One can find that 85.62\% of the total vehicles have PDRs exceeding 0.9 in the case of MCS index of 7, while, the MCS index of 11 suppresses the percentage 84.93\%, which are shown in Figs. \ref{fig_PDR}a and b, respectively}. It should also be noted that with an MCS index of 11, some vehicles witness a lower PDR dropping down to 0.3, whereas with an MCS index of 7, the lowest PDR observed is around 0.6.




\section{Conclusion}
\MK{
This paper has presented a comprehensive computer simulation framework that features ``vertical integration'' throughout geographic/traffic setup and PHY/RRC layers of C-V2X. The highlight of the contribution was the implementation of MCS, which completes a holistic view on the performance of a C-V2X network depending on the geographic/traffic.
}


\begin{figure}
\centering
\includegraphics[width = \linewidth]{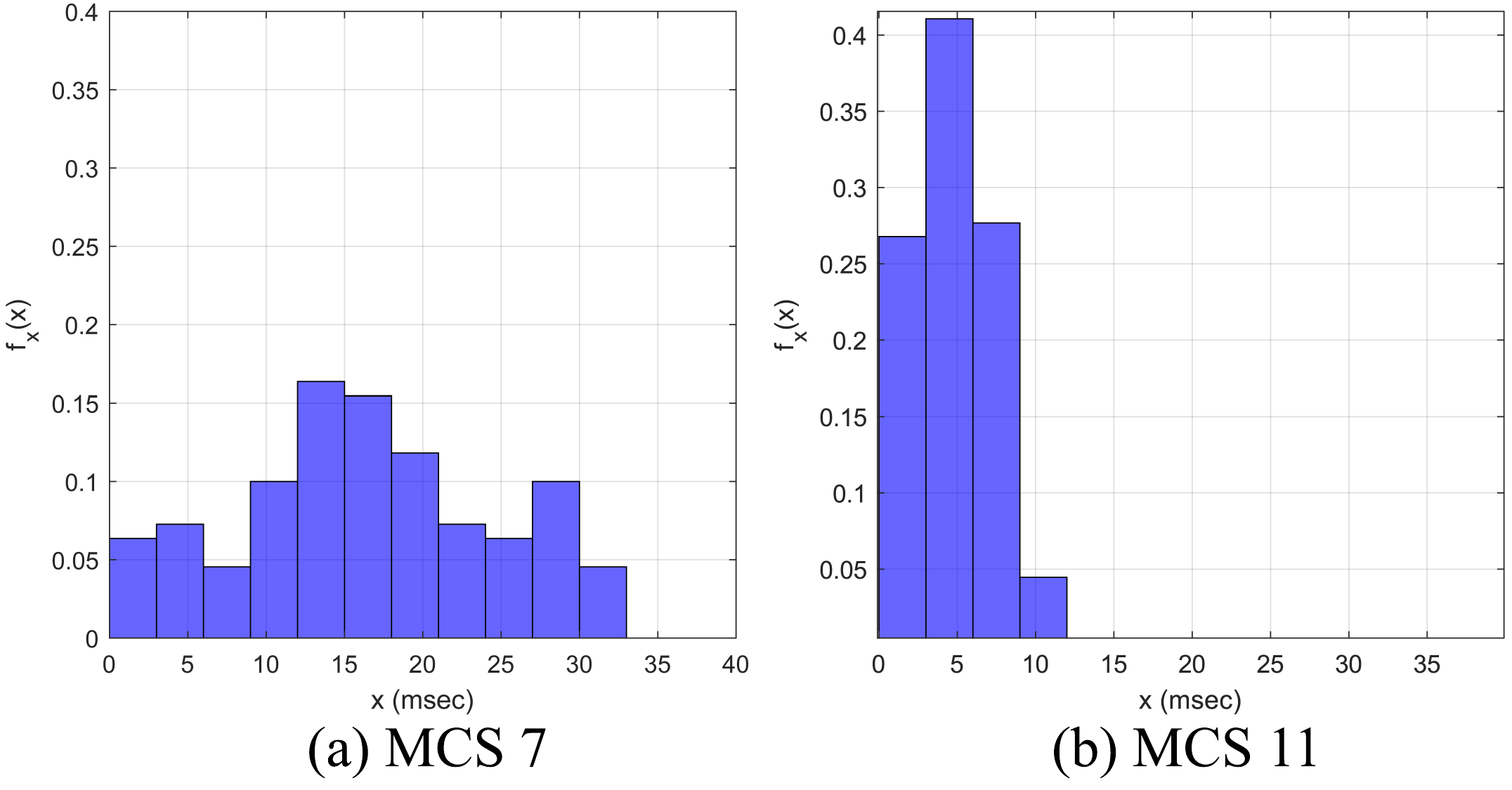}

\caption{PDF of latency}
\label{fig_latency}
 \end{figure}

\begin{figure}
\centering
\includegraphics[width = \linewidth]{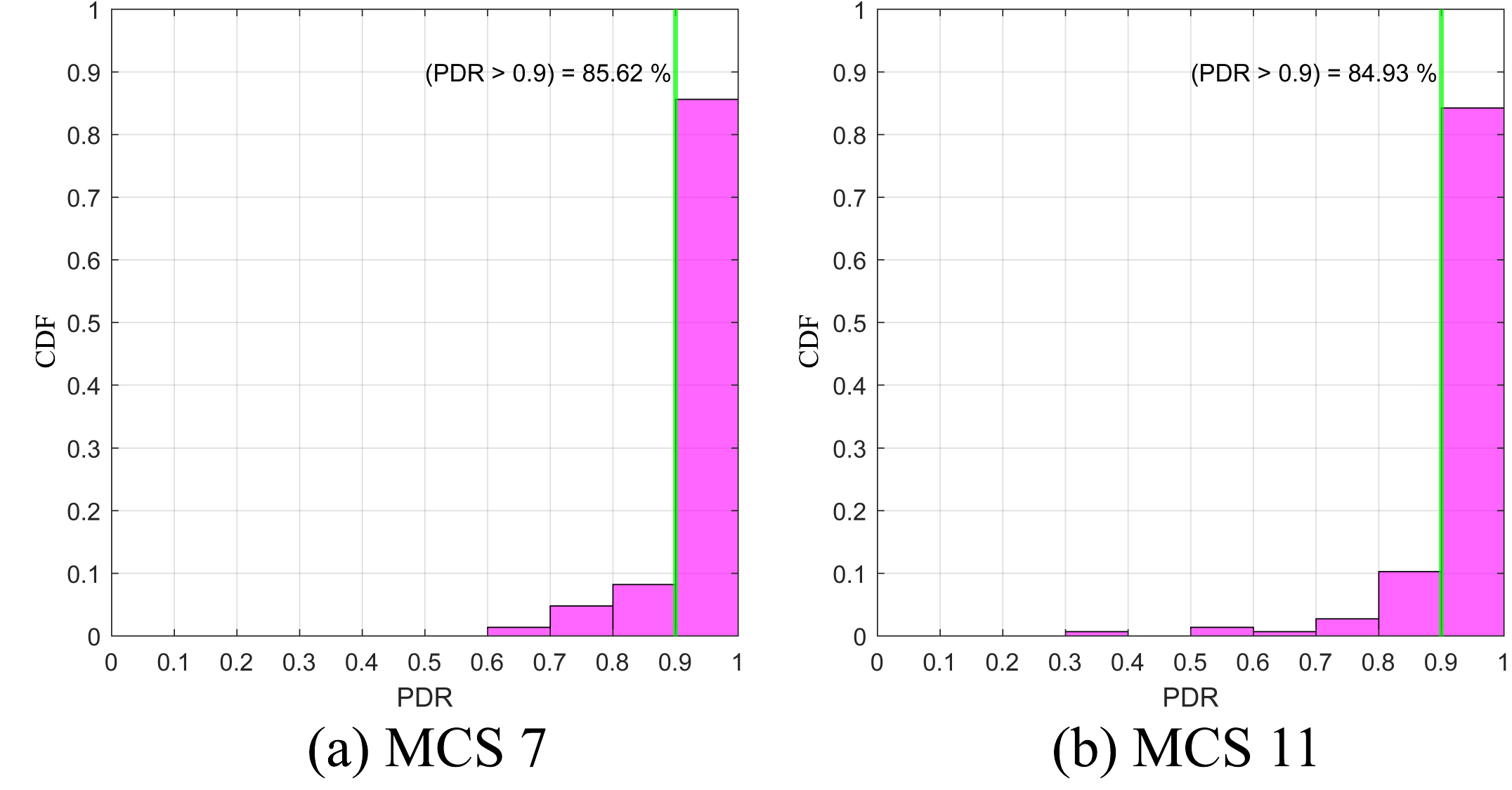}
\caption{CDF of PDR}
\label{fig_PDR}
 \end{figure}


\end{document}